\documentclass[aps,twocolumn,amsmath,letterpaper]{revtex4}
\usepackage{amssymb}
\usepackage{graphicx}
\usepackage{array}
\usepackage{hhline}
\usepackage{longtable}

\begin{document}

    \title{High $\textbf{q}$-State Clock Spin Glasses in Three Dimensions and the Lyapunov Exponents of Chaotic Phases and Chaotic Phase Boundaries}

    \author{Efe Ilker$^{1}$ and A. Nihat Berker$^{1,2}$}
 \affiliation{$^1$Faculty of Engineering and Natural Sciences, Sabanc\i~University, Tuzla 34956, Istanbul, Turkey,}
    \affiliation{$^2$Department of Physics, Massachusetts Institute of Technology, Cambridge, Massachusetts 02139, U.S.A.}

\begin{abstract}

Spin-glass phases and phase transitions for $q$-state clock models
and their $q \rightarrow \infty$ limit the XY model, in spatial
dimension $d = 3$, are studied by a detailed renormalization-group
study that is exact for the $d=3$ hierarchical lattice and
approximate for the cubic lattice. In addition to the now
well-established chaotic rescaling behavior of the spin-glass phase,
each of the two types of spin-glass phase boundaries displays, under
renormalization-group trajectories, their own distinctive chaotic
behavior.  These chaotic renormalization-group trajectories
subdivide into two categories, namely as strong-coupling chaos (in
the spin-glass phase and, distinctly, on the spinglass-ferromagnetic
phase boundary) and as intermediate-coupling chaos (on the
spinglass-paramagnetic phase boundary).  We thus characterize each
different phase and phase boundary exhibiting chaos by its distinct
Lyapunov exponent, which we calculate.  We show that, under
renormalization-group, chaotic trajectories and fixed distributions
are mechanistically and quantitatively equivalent. The phase
diagrams of arbitrary even $q$-state clock spin-glass models in
$d=3$ are calculated.  These models, for all non-infinite $q$, have
a finite-temperature spin-glass phase. Furthermore, the spin-glass
phases exhibit a universal ordering behavior, independent of $q$.
The spin-glass phases and the spinglass-paramagnetic phase
boundaries exhibit universal fixed distributions, chaotic
trajectories and Lyapunov exponents.  In the XY model limit, our
calculations indicate a zero-temperature spin-glass phase.




\end{abstract}

    \maketitle
    \def\s{\rule{0in}{0.28in}}
    \setlength{\LTcapwidth}{\columnwidth}

\section{Introduction}

Spin-glass phases, with randomly frozen local order
\cite{NishimoriBook} and chaotic behavior under scale change
\cite{McKayChaos,McKayChaos2,BerkerMcKay}, reflecting the effects of
frozen interaction disorder, competition, and frustration, remain a
uniquely fascinating and broadly relevant subject of statistical
mechanics and condensed matter physics. However, the large and
richly complex amount of theoretical knowledge produced on spin
glasses has been overwhelmingly derived from Ising, i.e., $s_i=\pm
1$, spin models.\cite{Heisenberg}

By contrast, we present here a detailed renormaliza-tion-group study
of spin-glass phases and phase transitions, for $q$-state clock
models and their $q \rightarrow \infty$ limit the XY model, in
spatial dimension $d = 3$.  We note that, in addition to the now
well-established chaotic behavior of the spin-glass phase \cite
{McKayChaos,McKayChaos2,BerkerMcKay,Bray,Hartford,Nifle1,Nifle2,Banavar,
Frzakala1,Frzakala2,Sasaki,Lukic,Ledoussal,Rizzo,Katzgraber,Yoshino,Pixley,
Aspelmeier1,Aspelmeier2,Mora,Aral,Chen,Jorg}, each of the two types
of spin-glass phase boundaries displays, under renormalization-group
trajectories, their own distinctive chaotic behavior.  We see that
these chaotic renormalization-group trajectories subdivide into two
categories, namely as strong-coupling chaos (in the spin-glass phase
and, distinctly, on the spinglass-ferromagnetic phase boundary) and
as intermediate-coupling chaos (on the spinglass-paramagnetic phase
boundary).  We thus quantitatively characterize each different phase
and phase boundary exhibiting chaos by its distinct Lyapunov
exponent as used in the general chaotic studies literature
\cite{Collet,Hilborn}, which we calculate. We show that, under
renormalization-group, chaotic trajectories and fixed distributions
are mechanistically and quantitatively equivalent.

We calculate and display the phase diagrams of arbitrary even
$q$-state clock spin-glass models in $d=3$.  These models, for any
non-infinite $q$, have a finite-temperature spin-glass phase.
Furthermore, we find that the spin-glass phases exhibit a universal
ordering behavior, independent of $q$.  The spin-glass phases and
the spinglass-paramagnetic phase boundary exhibit universal fixed
distributions, chaotic trajectories and Lyapunov exponents. In the
$d = 3$ XY model limit, our calculations indicate a zero-temperature
spin-glass phase.

\section{The $\textbf{q}$-state clock spin-glass model and the renormalization-group method}
The $q$-state clock models are composed of unit spins that are
confined to a plane and that can only point along $q$ angularly
equidistant directions. Accordingly, the $q$-state clock spin-glass
model is defined by the Hamiltonian
    \begin{equation}
        \label{eq:1}
        \begin{split}
            -\beta \mathcal{H}=&\sum_{\langle ij \rangle}
            J_{ij}\vec{s}_i\cdot\vec{s}_j = \sum_{\langle ij \rangle}
            J_{ij}cos(\theta_{i}-\theta_{j}),
        \end{split}
    \end{equation}
where $\beta=1/k_{B}T$, at site $i$ the spin angle $\theta_{i}$
takes on the values $(2\pi/q)\sigma_i$ with
$\sigma_i=0,1,2,...,q-1$, and $\langle ij \rangle$ denotes that the
sum runs over all nearest-neighbor pairs of sites.  The bond
strengths $J_{ij}$ are $+J>0$ (ferromagnetic) with probability $1-p$
and $-J$ (antiferromagnetic) with probability $p$. This model
becomes the Ising model for q=2 and the XY model for
$q\rightarrow\infty$.

\begin{figure}[h!]
\centering
\includegraphics[scale=1]{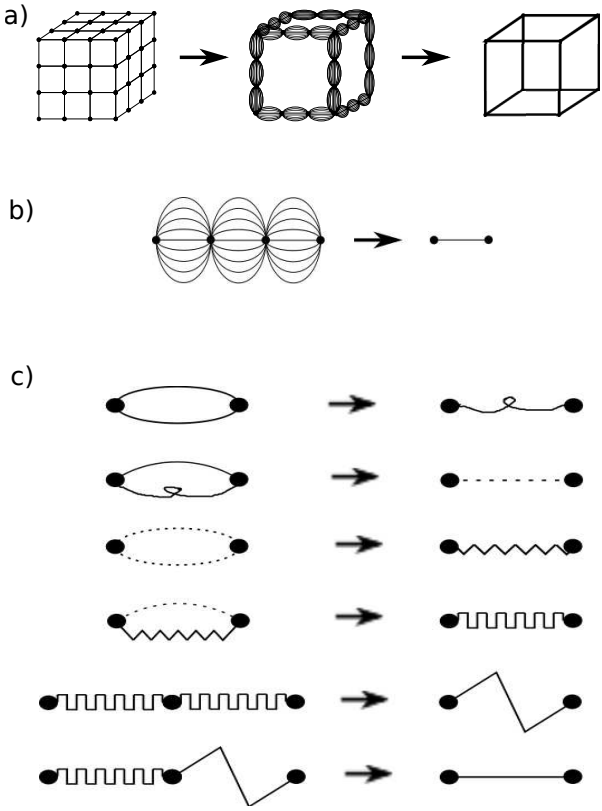}
\caption{(a) Migdal-Kadanoff approximate renormalization-group
transformation for the $d=3$ cubic lattice with the length-rescaling
factor of $b=3$. Bond-moving is followed by decimation. (b) Exact
renormalization-group transformation for the equivalent $d=3$
hierarchical lattice with the length-rescaling factor of $b=3$. (c)
Pairwise applications of the quenched probability convolution of
Eq.(5), leading to the exact transformation in (b).}\label{fig:1}
\end{figure}

The $q$-state clock spin-glass model, in $d=3$ dimensions, is
readily solved by a renormalization-group method that is approximate
on the cubic lattice \cite{Migdal,Kadanoff} and simultaneously exact
on the hierarchical lattice
\cite{BerkerOstlund,Kaufman1,Kaufman2,McKay,Hinczewski1}. Under
rescaling, for $q>4$, the form of the interaction as given in the
rightmost side of Eq.(1) is not conserved and one must therefore
express the Hamiltonian more generally, as
    \begin{equation}
        \label{eq:2}
        -\beta \mathcal{H}=\sum_{\langle ij
        \rangle}V(\theta_{i}-\theta_{j})\,.
    \end{equation}
Thus, the renormalization-group flows, for even $q$, are the flows
of $1+q/2$ interaction constants.  With no loss of generality, the
maximum value of $V(\theta_{i}-\theta_{j})$ is set to zero.

\begin{figure}[h!]
\centering
\includegraphics[scale=1]{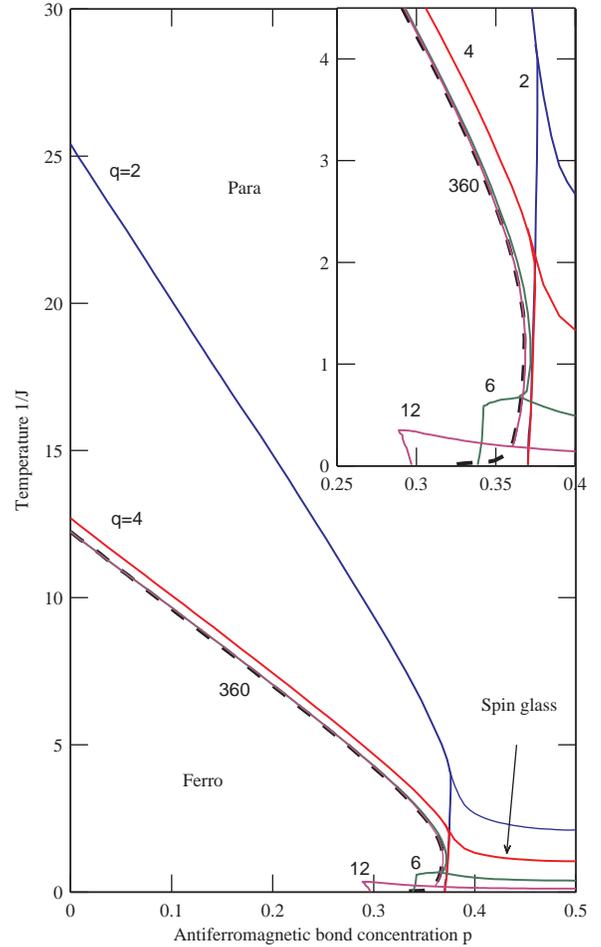}
\caption{(Color online) Calculated phase diagrams of the
$q=2,4,6,12$ clock spin-glass models in $d=3$ dimensions.  The phase
diagram for the XY limit, namely $q\rightarrow\infty$, is also
shown, by the dashed curve, calculated here with $q=360$ clock
states. As $q$ is increased, it is found that the spin-glass phase
retreats to lower temperatures while further protruding into the
ferromagnetic phase. In the XY limit, the spin-glass phase
disappears at zero temperature (see Fig. 4 below), as the phase
boundary between the remaining ferromagnetic and paramagnetic phases
numerically stabilizes, on the scale of the figure, for $q\gtrsim
6$.}\label{fig:1}
\end{figure}

The renormalization-group transformation, for spatial dimensions
$d=3$ and length rescaling factor $b = 3$ (necessary for treating
the ferromagnetic and antiferromagnetic correlations on equal
footing), is achieved by a sequence of bond moving
 \begin{equation}
        \label{eq:4}
        V_{bm}(\theta_1-\theta_2) + G_{12} = \sum_{n=1}^{b^{d-1}}
        V_n(\theta_1-\theta_2)
    \end{equation}
and decimation
\begin{equation}
        \label{eq:3}
        e^{V_{dec}(\theta_1-\theta_4)+G_{14}}=\sum_{\theta_2,\theta_3}
        e^{V_1(\theta_1-\theta_2)+V_2(\theta_2-\theta_3)+V_3(\theta_3-\theta_4)},
    \end{equation}
where the constants $G_{ij}$ are fixed by the requirement that the
maximum value of $V(\theta_{i}-\theta_{j})$ is zero.

\begin{figure}[h!]
\centering
\includegraphics[scale=1]{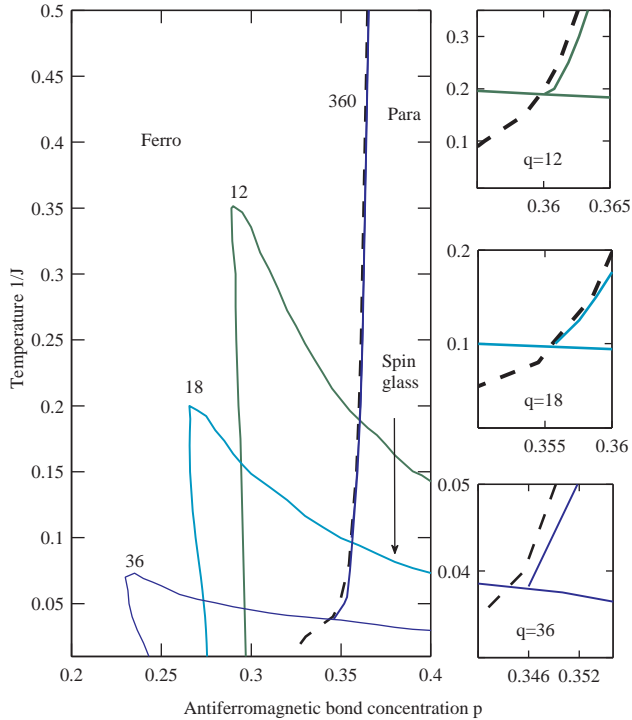}
\caption{(Color online) Phase diagrams of the $q=12,18,36$ clock
spin-glass models in $d=3$ dimensions. Two new phenomena are
simultaneously detected here: (1) double reentrance:
paramagnetic-ferromagnetic-spinglass-ferromagnetic as temperature is
lowered, (2) lateral reentrance:
ferromagnetic-spinglass-ferromagnetic-paramagnetic as p is
increased. The XY limit is given, calculated here with $q=360$ clock
states, by the dashed curve. The panels on the right show the region
of the multicritical point, where the three phase boundary lines
meet, for each $q$ case along with the ferromagnetic-paramagnetic
phase boundary of the XY limit.}\label{fig:2}
\end{figure}

\begin{figure}[h!]
\centering
\includegraphics[scale=1]{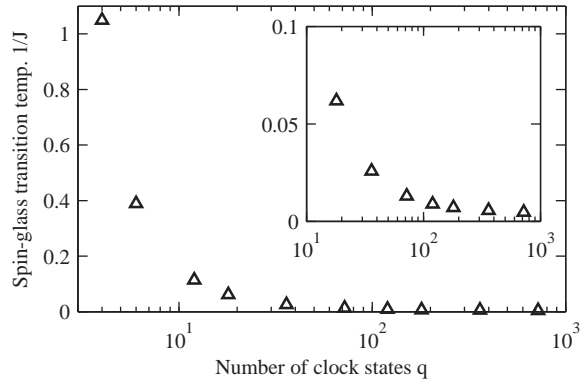}
\caption{The calculated transition temperatures between the
spin-glass phase and the paramagnetic phase, at $p = 0.5$, as a
function of $q$, up to very large values of $q = 720$. Note the
logarithmic scale of the horizontal axes.  The slow decay of the
transition temperature suggests that a zero-temperature spin-glass
phase exists in the $(q\rightarrow\infty)$ XY model
limit.}\label{fig:1}
\end{figure}

The starting bimodal quenched probability distribution of the
interactions, characterized by $p$ and described above, is also not
conserved under rescaling. The renormalized quenched probability
distribution of the interactions is obtained by the convolution
\cite{Andelman}
\begin{multline}
\label{eq:5}
        P'(V'(\theta_{i'j'})) =\\
        \int{\left[\prod_{ij}^{i'j'}dV(\theta_{ij})P(V(\theta_{ij}))\right]}
         \delta(V'(\theta_{i'j'})-R(\left\{V(\theta_{ij})\right\})),
\end{multline}
where $R(\left\{V(\theta_{ij})\right\})$ represents the bond moving
and decimation given in Eqs.(3) and (4).  For numerical
practicality, the bond moving and decimation of Eqs.(3) and (4) are
achieved by a sequence of pairwise combination of interactions, as
shown in Fig.1(c), each pairwise combination leading to an
intermediate probability distribution resulting from a pairwise
convolution as in Eq.(5). We effect this procedure numerically, by
generating 5,000 interactions that embody the quenched probability
distribution resulting from each pairwise combination.  Each of the
generated 5,000 interactions is determined by $1+q/2$ interaction
constants.  At each pairwise convolution as in Eq.(5), 5,000
randomly chosen pairs are matched by Eq.(3) or (4), and a new set of
5,000 is produced.  We have checked that our results are insensitive
to further increasing the number 5,000. Furthermore, our calculated
phase diagrams exactly match, for $q=2$, the results in
Refs.\cite{Migliorini,Hinczewski,Guven,Gulpinar} which are
numerically exact by the use of the histogram representation of the
quenched probability distribution.

The different thermodynamic phases of the model are identified by
the different asymptotic renormalization-group flows of the quenched
probability distributions.  For all renormalization-group flows,
inside the phases and on the phase boundaries, Eq.(5) is iterated
until asymptotic behavior is reached, meaning that we are studying
an effectively infinite hierarchical lattice. Thus, we are able to
calculate phase diagrams for any number of clock states $q$. Our
results are obtained by averaging over 30 to 50 different
realizations of the initial $\pm J cos(\theta_{i}-\theta_{j})$
distribution into the 5,000 initial interactions.  In this study, we
consider even values of $q$ and the calculated phase diagrams are
symmetric around $p = 0.5$ with the antiferromagnetic phase
replacing the ferromagnetic phase, so that only the $p=0$ to 0.5
halves are shown below.  If $q$ is odd, the system does not have
sublattice spin-reversal symmetry, which leads to asymmetric phase
diagrams.

\section{Calculated phase diagrams for $d=3$ $\textbf{q}$-state clock and XY spin glasses}

\begin{figure}[h!]
\centering
\includegraphics[scale=1]{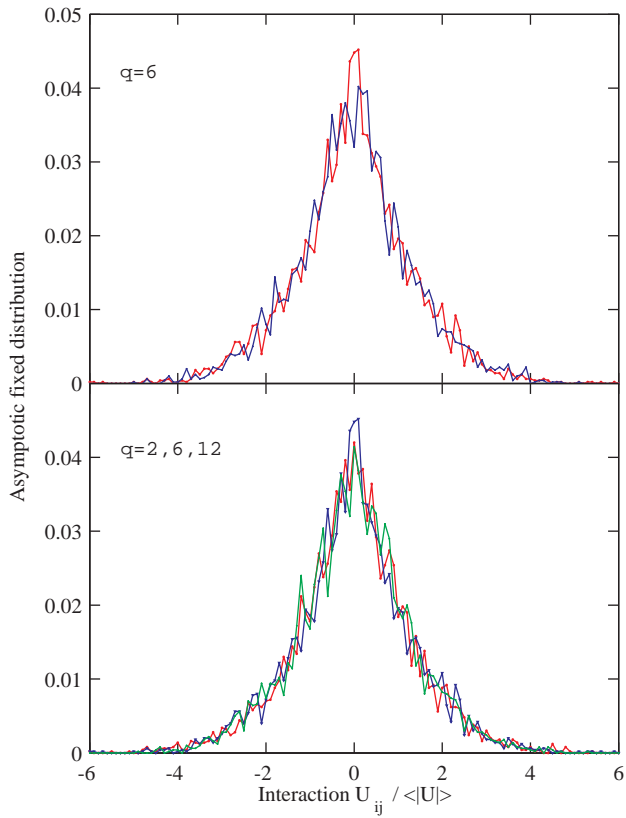}
\caption{(Color online) Asymptotic fixed distribution, under
renormalization-group transformations, of the interactions in the
spin-glass phase, namely the renormalization-group sink of the
spin-glass phase.  Note that these are strong-coupling
distributions, as the average interaction strength $<|U|>$ diverges
to infinity under the renormalization-group transformations.  The
divergence of $<|U|>$ is as $b^{0.24 n}$, where $n$ is the number of
iterations.  Top: For the $q = 6$-state clock model in $d=3$, for a
trajectory starting in the spin-glass phase at $p=0.5$ and
temperature $1/J = 0.05$, the distributions after 20 and 21
renormalization-group steps are shown. It is seen that these two
distributions coincide, signifying a fixed distribution. Bottom:
Asymptotic fixed distributions of the spin-glass phases for the $q =
2$ (Ising), $6,12$-state clock models in $d=3$. These distributions
are reached after 20 renormalization-group steps, starting at
$p=0.5$ and temperature $1/J = 0.05$.  Note that the spin-glass sink
fixed distributions for different values of $q$ coincide. The
Lyapunov exponent is $\lambda = 1.93$ for the single, universal
distribution that is illustrated in this figure.}
\end{figure}

\begin{figure}[h!]
\centering
\includegraphics[scale=1]{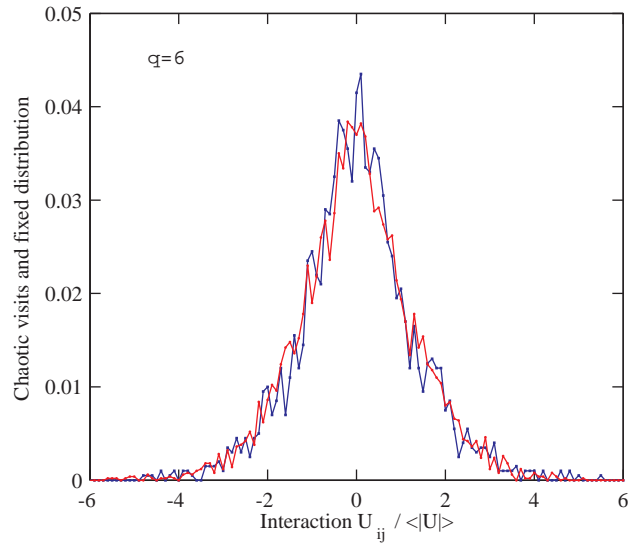}
\caption{(Color online) Comparison, showing the coincidence, of the
chaotic visits of the consecutively renormalized interactions at a
given position of the system (for 2,000 renormalization-group
iterations) and of the asymptotic distribution of the interactions
across the system at a given renormalization-group step, for the
spin-glass phase.  The Lyapunov exponent is $\lambda = 1.93$.
Identical asymptotic behavior occurs for all $q$-state clock
spin-glass phases, as shown in Fig.5.  The divergence of $<|U|>$ is
as $b^{0.24 n}$, where $n$ is the number of iterations. }
\end{figure}

Our calculated phase diagrams for the $q=2,4,6,12$ clock spin-glass
models are shown together in Fig. 2.  The phase diagram for the XY
limit, namely $q\rightarrow\infty$, is also shown in Fig. 2,
calculated here with $q=360$ clock states. In this limit, the
spin-glass phase disappears at zero temperature, whereas the phase
boundary between the ferromagnetic and paramagnetic phases
numerically stabilizes, on the scale of the figure, for $q\gtrsim
6$. The paramagnetic-ferromagnetic-spinglass reentrance as
temperature is lowered, previously seen \cite{Migliorini,Roy} for
$q=2$, namely the Ising case, is also seen here for the other $q$.
As $q$ is increased, it is found that the spin-glass phase retreats
to lower temperatures while further protruding into the
ferromagnetic phase.

The calculated phase diagrams for the high-$q$ models, $q =
12,18,36,360$, are shown in Fig 3.  As $q$ is increased, the trend
mentioned above, of the spin-glass phase retreating to lower
temperatures while further protruding into the ferromagnetic phase,
is also seen here.  Furthermore, two new phenomena are
simultaneously detected here: (1) double reentrance, namely
paramagnetic-ferromagnetic-spinglass-ferromagnetic phases as
temperature is lowered; (2) lateral reentrance, namely
ferromagnetic-spinglass-ferromagnetic-paramagnetic phases as p is
increased.  Multiple reentrances have previously been seen in liquid
crystal systems.\cite{Indekeu,Netz,Mazza}

The (slow) disappearance of the $q$-state clock spin-glass phase is
shown in Fig. 4, where the calculated spin-glass transition
temperatures at $p = 0.5$ are shown as a function of $q$, up to very
large values of $q = 720$. The slow decay of the transition
temperature suggests that a zero-temperature spin-glass phase
\cite{Grinstein} exists in the $q\rightarrow\infty$, namely XY model
limit, in agreement with the previous Monte Carlo study of
Ref.\cite{Pixley}.

\begin{figure}[h!]
\centering
\includegraphics[scale=1]{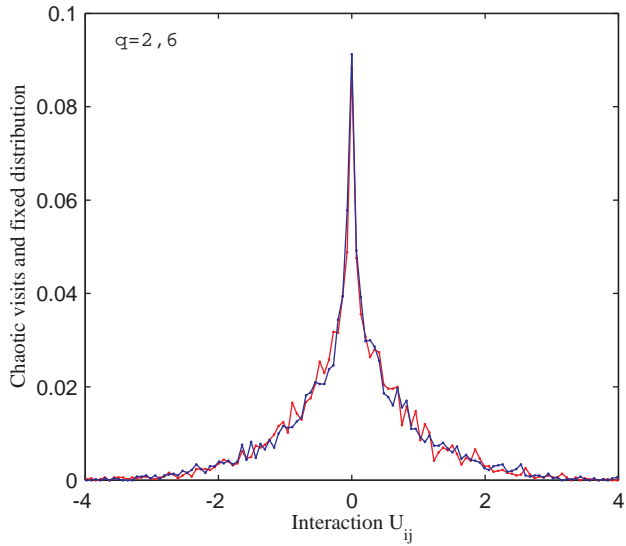}
\caption{(Color online) The fixed distribution and, equivalently,
chaotic renormalization-group trajectory onto which the phase
boundary between the spin-glass and paramagnetic phases
renormalizes, for the $q=2$ and $q=6$-state clock models in $d=3$.
The Lyapunov exponent is $\lambda = 1.35$. Identical asymptotic
behavior occurs for all $q$.  Note that the chaotic behavior and
fixed distribution are at intermediate coupling strength, with
$<|U|> = 0.686$ for all $q$.}
\end{figure}

\section{Stable Fixed Distribution and Chaotic Renormalization-Group Trajectory of Clock Spin-Glass Phases}

\subsection{Stable Fixed Distribution}

For the spin-glass phase of the Ising model ($q=2$), under repeated
renormalization-group transformations, the quenched probability
distribution of the interactions across the system becomes symmetric
in ferromagnetic ($J_{ij}>0$) and antiferromagnetic ($J_{ij}<0$)
couplings, with the average magnitude of either type of interaction
equal and diverging to infinity.\cite{Migliorini}

\begin{figure}[h!]
\centering
\includegraphics[scale=1]{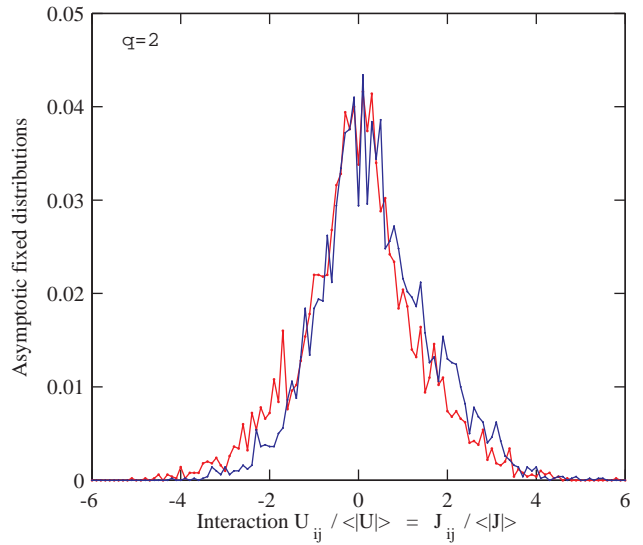}
\caption{(Color online) Two different, non-coinciding
strong-coupling fixed distributions: The fixed distribution and,
equivalently, chaotic renormalization-group trajectory onto which
the spinglass-ferromagnetic phase boundary of the $q = 2$-state
model in $d=3$ renormalizes. For comparison and distinction, the
asymptotic fixed distribution and chaotic renormalization-group
trajectory of the corresponding spin-glass phase is also shown.  The
latter curve is symmetric around $U_{ij}/<|U|> = 0$, whereas the
former curve is noticeably displaced towards positive
(ferromagnetic) interaction values. Note that these are
strong-coupling distributions and are therefore shown as a fraction
of the diverging $<|U|>$.  The divergence of $<|U|>$ is as $b^{0.46
n}$ and $b^{0.24 n}$, where $n$ is the number of iterations,
respectively for the phase boundary and phase sink cases. Thus, even
small shifts in the shown curves signify large differences in the
interaction values. The Lyapunov exponents are $\lambda = 1.69$ and
1.93, respectively for the phase boundary and phase sink cases. For
$q=2$, the form of the interaction in Eq.(1) is conserved under
renormalization, which is reflected in the horizontal axis label
here.}
\end{figure}

For the spin-glass phases of all $q$-state clock models, we find
that under repeated renormalization-group transformations, the
interaction values $V_{ij}(\theta)$ divide into two groups:
$\theta=2\pi n/q$ and $\theta=2\pi m/q$, where n is an even integer,
n=0,2,4,...,q-2, and m is an odd integer, m=1,3,5,...,q-1. The
asymptotic renormalized quenched distribution of the interactions is
symmetric, with interactions equal within each group mentioned
above, but at each location overwhelmingly favoring one or the other
of the two groups. Thus, the interaction difference between the two
groups,
\begin{equation}
\frac{1}{q/2}\sum_{n=0,2,...}^{q-2} V_{ij}(2\pi n/q) -
\frac{1}{q/2}\sum_{m=1,3,...}^{q-1} V_{ij}(2\pi m/q) \, \equiv \,
U_{ij},
\end{equation}
after many renormalization-group transformations, is randomly and
equally distributed as positive or negative in our sampling of 5,000
interactions, which represent the distribution of interactions
spatially across the system, with the average magnitude of either
type of interaction equal and diverging to infinity as $b^{0.24 n}$,
where $n$ is the number of iterations.  This asymptotic fixed
distribution, namely the sink of the spin-glass phase, is shown in
Fig. 5.  Note that this asymptotic behavior is also consistent with
the behavior of the Ising spin-glass phase ($U$ reduces to $J$ for
the Ising case), recalled at the beginning of this section. This
behavior leaves the system asymptotically frustrated.

\subsection{Chaotic Renormalization-Group Trajectory}

The fact that the Ising spin-glass phase is characterized by the
chaotic rescaling behavior of the interactions
\cite{McKayChaos,McKayChaos2,BerkerMcKay}, and therefore of the
correlations \cite{Aral}, is now well established \cite
{McKayChaos,McKayChaos2,BerkerMcKay,Bray,Hartford,
Nifle1,Nifle2,Banavar,Frzakala1,Frzakala2,Sasaki,Lukic,Ledoussal,Rizzo,
Katzgraber,Yoshino,Pixley,Aspelmeier1,Aspelmeier2,Mora,Aral,Chen,Jorg}
and is also seen here for the spin-glass phases of the $q$-state
clock models in $d = 3$, as shown in Fig. 5.  As with the Ising
model \cite{Hartford}, we have here a strong-coupling chaotic
behavior: The values of the interaction difference $U_{ij}$ obtained
by successive renormalization-group transformations at any specific
location, divided by the average magnitude $<|U|>$ across the
system, fall into a chaotic band. Thus, the $U_{ij}/<|U|>$ values
are sampled within the band as shown in Fig. 5.  The average
magnitude $<|U|>$ diverges to infinity under repeated
renormalization-group transformations, as $b^{0.24 n}$, where $n$ is
the number of iterations.

We thus realize that the spin-glass phases can be characterized by
the Lyapunov exponent of general chaotic behavior
\cite{Collet,Hilborn}.  The positivity of the Lyapunov exponent
measures the strength of the chaos \cite{Collet,Hilborn} and was
also used in the previous spin-glass study of Ref.[23].  The
calculation of the Lyapunov exponent is applied here to the chaotic
renormalization-group trajectory at any specific location in the
lattice,
\begin{equation}
\lambda = \lim _{n\rightarrow\infty} \frac{1}{n} \sum_{k=0}^{n-1}
\ln \Big|\frac {dx_{k+1}}{dx_k}\Big|
\end{equation}
where $x_k = U_{ij}/<|U|>$ at step $k$ of the renormaliza-tion-group
trajectory.  The sum in Eq.(7) is to be taken within the asymptotic
chaotic band.  Thus, we throw out the first 100
renormalization-group iterations to eliminate the points outside of,
but leading to the chaotic band. Subsequently, typically using 2,000
renormalization-group iterations in the sum in Eq.(7) assures the
convergence of the Lyapunov exponent value.  We have calculated the
Lyapunov exponent $\lambda = 1.93$ for the clock spin-glass phases
of $q = 2,6,12$ and presumably for the clock spin-glass phases of
all $q$, which is to be expected since all $q$ spin-glass phases
renormalize to the same chaotic band, as seen in Fig.5.

\subsection{Equivalence of the Chaotic Renormalization-Group Trajectory and the Quenched Probability Fixed Distribution}

\begin{figure}[h!]
\centering
\includegraphics[scale=1]{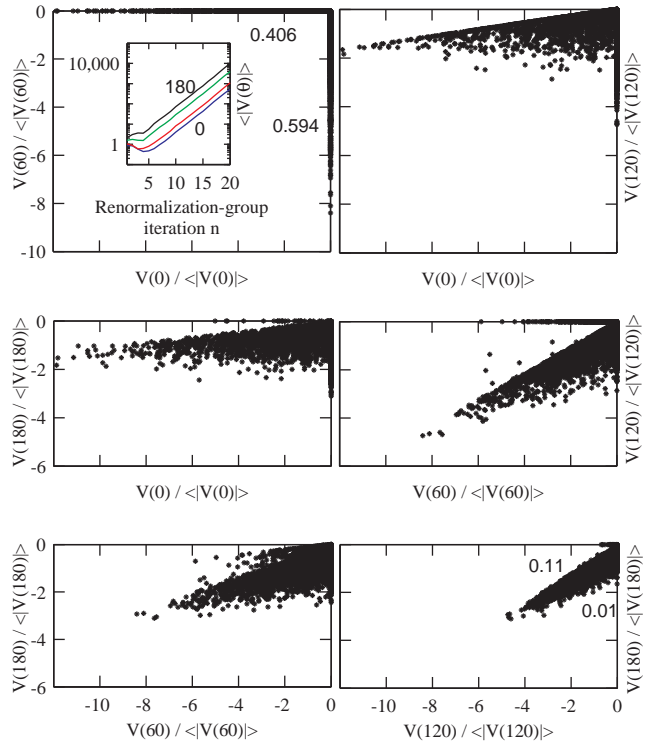}
\caption{The fixed distribution and, equivalently, chaotic
renormalization-group trajectory onto which the
spinglass-ferromagnetic phase boundary of the $q = 6$-state clock
model in $d=3$ renormalizes. The coupled distributions of the
interactions $V(\theta)/<|V(\theta)|>$ are shown, while next to each
axis, the fraction of points on the axis are given. The full fixed
distribution has the form $P(\{V(0),V(60), V(120), V(180)\})$, as a
coupled function of its arguments. This is a strong-coupling
behavior: The inset in the upper left panel shows, with the
logarithmic vertical scale, the diverging $<|V(\theta)|>$ as a
function of renormalization-group iteration $n$, the consecutive
curves being for $\theta = 180, 120, 60, 0$.  It is seen that, for
all $\theta$, $<|V(\theta)|>$ diverges as $b^{0.46 n}$, where $n$ is
the number of iterations.}
\end{figure}

The distributions of the interaction difference $U_{ij}$ values
shown in Figs. 5 and 6, respectively obtained as the spatial
distribution across the system after many renormalization-group
transformations and the values obtained by successive
renormalization-group transformations at a specific location in the
system, are in fact identical, as seen in Fig. 6.  Thus, it is
understood that the asymptotic fixed distribution is realized, after
a given number of renormalization-group transformations, by the
interactions at different locations being at different points of the
same chaotic trajectory.

\section{Unstable Fixed Distributions and Chaotic Renormalization-Group Trajectories of the Clock Spinglass-Paramagnetic and Spinglass-Ferromagnetic Boundaries}

We find that the points on the various spin-glass phase boundaries
also renormalize to a fixed distribution of the quenched
interactions across the system and, equivalently, to a chaotic
renormalization-group trajectory of the interaction at any single
location in the lattice.  The difference between the asymptotic
rescaling behaviors inside the spin-glass phase and on the
spin-glass phase boundaries is that, under rescaling
transformations, the fixed distribution and the chaotic trajectory
are reached in a stable manner, with respect to initial conditions,
for the spin-glass phase and are conversely unstable for the
spin-glass phase boundaries.  The ferromagnetic-paramagnetic phase
boundary renormalizes to the pure ferromagnetic system, where an
unstable fixed point determines the critical exponent, differently
for each $q$.

\subsection{The Spinglass-Paramagnetic Phase Boundary}

The phase boundary between the spin-glass and paramagnetic phases
renormalizes to the fixed distribution and chaotic
renormalization-group trajectory shown in Fig. 7.  The interaction
grouping, under rescaling, described before Eq.(6) also happens.
However, this behavior here occurs at finite coupling $<|U|> =
0.686$ for all $q$, in contrast to the asymptotic behaviors of the
spin-glass phase (given above) and of the spinglass-ferromagnetic
phase boundary (given below), which occur at strong coupling  $<|U|>
\rightarrow \infty$.

\subsection{The Spinglass-Ferromagnetic Phase Boundary}

The phase boundary between the spin-glass and ferromagnetic phases
renormalizes to a fixed distribution and chaotic
renormalization-group trajectory at strong coupling $<|U|>
\rightarrow \infty$.  The interaction grouping, under rescaling,
described before Eq.(6) does not happen.  Thus, the interaction
$V(\theta)$ as a function of $\theta$ has $1 + q/2$ different
values. The asymptotic fixed distribution for $q = 2$ is shown in
Fig. 8 and is characterized by the Lyapunov exponent $\lambda =
1.69$.  The divergence of $<|U|>$ is as $b^{0.46 n}$, where $n$ is
the number of iterations.

\begin{table}[h!]
\begin{tabular}{c c c c c c c c c c c c c c c c}
\hline
Weight in  &\vline & &\vline &    &\vline &   &\vline &  \\
fixed dist.  &\vline & $e^{V(0)} $ &\vline &  $e^{V(60)}$  &\vline &  $e^{V(120)}$ &\vline &  $e^{V(180)}$\\
\hline
0.4802  &\vline & 1 &\vline &  0  &\vline &  0 &\vline &  0\\
\hline
0.3951  &\vline & 0 &\vline &  1  &\vline &  0 &\vline &  0\\
\hline
0.1139  &\vline & 1 &\vline &  0  &\vline &  1/2 &\vline &  0\\
\hline
0.0108  &\vline & 0 &\vline &  1  &\vline &  0 &\vline &  2/3\\
\hline \hline
\end{tabular}
\caption{Dominant potentials in the asymptotic fixed distribution of
the phase boundary between the spin-glass and ferromagnetic phases
of the $q = 6$ clock model in $d = 3$.  Thus, the system remains
frustrated at all length scales.\\} \label{tab:1}
\end{table}

The asymptotic fixed distribution for $q = 6$ is shown in Fig. 9.
The full fixed distribution has the form $P(\{V(0),V(60), V(120),
V(180)\})$, as a coupled function of its arguments.  The dominant
configurations of this fixed distribution are shown in Table I.  The
system remains frustrated at all length scales.

\section{Conclusion}

We have calculated, from renormalization-group theory, the phase
diagrams of arbitrary even $q$-state clock spin-glass models in
$d=3$.  These models, for all non-infinite $q$, have a
finite-temperature spin-glass phase, exhibiting a universal ordering
behavior, independent of $q$. In addition to the chaotic rescaling
behavior of the spin-glass phase, each of the two types of
spin-glass phase boundaries displays, under renormalization-group
trajectories, their own distinctive chaotic behavior, subdividing
into two categories: strong-coupling chaos, in the spin-glass phase
and distinctly on the spinglass-ferromagnetic phase boundary, and
intermediate-coupling chaos, on the spinglass-paramagnetic phase
boundary. We uniquely characterize each different phase and phase
boundary exhibiting chaos by its distinct Lyapunov exponent from
general chaos studies, which we calculate. We demonstrate that,
under renormalization-group, chaotic trajectories and fixed
distributions are mechanistically and quantitatively equivalent. The
spin-glass phases and the spinglass-paramagnetic phase boundaries
exhibit universal fixed distributions, chaotic trajectories and
Lyapunov exponents.  In the XY model limit, our calculations
indicate a zero-temperature spin-glass phase.

\begin{acknowledgments}
Support by the Alexander von Humboldt Foundation, the Scientific and
Technological Research Council of Turkey (T\"UBITAK), and the
Academy of Sciences of Turkey (T\"UBA) is gratefully acknowledged.
\end{acknowledgments}

\end{document}